\documentclass[twocolumn]{aastex631}

\usepackage{amsmath} 
\usepackage{soul}

\shorttitle{Inflated Eccentric Migration of Evolving Gas-Giants II}
\shortauthors{Glanz et al.}

\graphicspath{{./}{figures/}}

\begin{document}

\title{Inflated Eccentric Migration of evolving gas giants II: Numerical methodology and basic concepts}

\correspondingauthor{Hila Glanz}
\email{Glanz@tx.technion.ac.il}

\author[0000-0002-6012-2136]{Hila Glanz}
\affiliation{Technion - Israel Institute of Technology, Haifa, 3200002, Israel}

\author[0000-0002-2728-0132]{Mor Rozner}
\affiliation{Technion - Israel Institute of Technology, Haifa, 3200002, Israel}


\author[0000-0002-5004-199X]{Hagai B. Perets}
\affiliation{Technion - Israel Institute of Technology, Haifa, 3200002, Israel}

\author[0000-0001-7113-723X]{Evgeni Grishin}
\affiliation{Technion - Israel Institute of Technology, Haifa, 3200002, Israel}
\affiliation{School of Physics and Astronomy, Monash University, Clayton 3800, VIC, Australia}

\begin{abstract}
Hot and Warm Jupiters (HJs\&WJs) are gas-giant planets orbiting their host stars at short orbital periods, posing a challenge to their efficient in-situ formation. Therefore, most of the HJs\&WJs are thought to have migrated from an initially farther-out birth locations. Current migration models, i.e disc-migration (gas-dissipation driven) and eccentric-migration (tidal evolution driven), fail to produce the occurrence rate and orbital properties of HJs\&WJs. 
Here we study the role of the thermal evolution and its coupling to tidal evolution.  We use the \texttt{AMUSE}, numerical environment, and \texttt{MESA}, planetary evolution modeling, to model in detail the coupled internal and orbital evolution of gas-giants during their eccentric-migration. In a companion paper, we use a simple semi-analytic model, validated by our numerical model, and run a population-synthesis study. 
We consider the initially inflated radii of gas-giants (expected following their formation), as well study the effects of  the potential slowed contraction and even re-inflation of gas-giants (due to tidal and radiative heating) on the eccentric-migration. Tidal forces that drive eccentric-migration are highly sensitive to the planetary structure and radius. Consequently, we find that this form of inflated eccentric-migration operates on significantly (up to an order of magnitude) shorter timescales than previously studied eccentric-migration models. Thereby, inflated eccentric-migration gives rise to more rapid formation of HJs\&WJs, higher occurrence rates of WJs, and higher rates of tidal disruptions, compared with previous eccentric migration models which consider constant $\sim$Jupiter radii for HJ\&WJ progenitors. Coupled thermal-dynamical evolution of eccentric gas-giants can therefore play a key-role in their evolution. 
  
\end{abstract}

\section{Introduction} \label{sec:intro}

Gas giant planets are thought to have formed from either core-accretion, in which runaway gas accretion takes place onto the massive, $\sim 10 M_\earth$, core \citep{Perri1974,BodenheimerPollack1986}, or from a direct collapse from the gas disk \citep{Mizuno1980,Boss1997,Armitage2010}. However, as the efficiencies of both channels are greatly affected by the local environment properties, such as the temperature, density, composition and velocities, they cannot solely describe the formation of gas giants that have extremely short period orbits around their host stars. These include the population of Hot Jupiters (HJs), with orbits of a few days \citep{Bodenheimer2000,Rafikov2005}, as well as some Warm Jupiters (WJs) with small pericenters.  The remaining non-negligible fraction of more distant WJs, might still be formed in-situ, as discussed in \citealt{HuangWuTriad2016,AndersonLaiPu2020}).
 Consequently, HJs\&WJs are thought to have formed at larger separations from their stars, and migrated inwards due to dynamical interactions, either with other bodies leading to high eccentricity migration, or with the gas from the protoplanetary disk, producing a drag force (see \citealt{Dawson2018} for a review). Nevertheless, past studies on these migration models could not reproduce the observed formation rates and properties of the current population of HJs\&WJs \citep{Dawson2018,ZhuDong2021}, as the typical migration timescales are potentially too long as to produce the inferred numbers of HJs/WJs and their appropriate timescales.

A planet orbiting its host star at close separation experiences significant tidal forces raised by the host star. The gravitational interaction between star and the bulge raised due to tides on the planet (and to a much lesser degree the tides raised on the star by the planet), eventually give rise to dissipation of orbital energy in the planetary atmosphere. This, in turn, leads to the orbital decay of the planet into shorter periods and more circular orbits. Consequently, planets on highly eccentric orbits with close peri-center approach to the host star may experience tidal migration, generally termed eccentric migration. 

The strength of the tides strongly depends on the planetary radius, which is typically considered as some constant $\sim$ Jupiter radius, R$_J$ in eccentric migration models. However, gas-giants are thought to form with far larger inflated radii and then cool and contract, where external heating by radiation and/or tidal heating may slow down their cooling and possibly even re-inflate them. 
After reaching large radii of up to $10 \  R_J$ by the end of the core accretion \citep{GinzburgChiang2019}, the gas giants contract to smaller radii, initially in a rapid process to $~4 \ R_J$ \citep{Guillot1996}, followed by a slower thermal contraction within a Kelvin-Helmholtz timescale  ($\sim {10}^8\  \rm{yrs}$), reaching radii of $\sim1.5-2.5R_J$ that continuously shrink in an even slower rate, depending on their mass and external energies. 
Since tidal migration depends strongly on the planetary radius, inflated planets could give rise for far faster eccentric migration compared with non-evolving constant Jupiter radii gas-giants typically considered in eccentric migration models. 
As the radius of such planets might decrease in a comparable timescale to the high eccentricity migration timescale, considering the internal evolution of the planet (initially a thermal contraction and cooling), can therefore play a key role in their dynamical evolution.

Here and in a companion paper (\citealt{Rozner2021}; hereafter paper I) we explore for the first time a self-consistent thermal-dynamical evolution of migrating planets over a wide parameter space, and throughout their evolution beginning at very high eccentricities (but see \citealp{Wu2006,MillerFortneyJackson2009,Petrovich2015a} where some of these issues were partially studied). 
We couple between the thermal evolution of gas-giants, and their dynamical evolution through eccentric tidal migration, as well as consider possible re-inflation and slow contraction of the planets due to external heating sources. 
 We find that eccentric migration of such inflated Jovian planets, which we term \emph{inflated eccentric migration}, significantly alters their dynamical evolution and could play a key role  in any type of eccentric migration, and in particular give rise to much (up to an order of magnitude) faster eccentric migration.

Here we present our numerical method, where we use \texttt{MESA} and \texttt{AMUSE} to accurately simulate the internal evolution of these planets during their migration. Our numerical results can be used to study the effect of other types of dynamical evolution and external energy sources. In Paper I we present a semi-analytical approach to simulate such migration, where we use the same equations of motion, but a simple modeling of the internal/thermal evolution, which therefore requires less computational resources in order to be used. Here we present some comparisons between the results of both methods, and find a good agreement. This also validates our use of the semi-analytical approach in the study and characterization of a large population of HJs\&WJs progenitors population, which we present in paper I. 

In the next section we describe our calculation method. We first discuss the considered external energy sources affecting the evolution of the giant planets (Sec. \ref{subsec:external_energies}), then we explain the mechanism of the high eccentricity migration with the different tide models in Subsec. \ref{sec: high eccentricity tidal migration}. Later (Subsec. \ref{sec:numerical}), we  describe our numerical simulation method to couple the dynamical and tidal evolution of the planets with their thermal evolution. In Sec. \ref{sec:Results} we present our results and their implications on the formation of HJs\&WJs, followed by discussions in Sec. \ref{Discussion}, and finally we summarize in Sec. \ref{sec:Summary}.

\section{Methods}
\label{sec:Methods}

\subsection{External energy sources}
\label{subsec:external_energies}
With the absence of any internal heating sources, following its formation and final runaway accretion stages, a newly-born gas giant begins to continuously cool-down and contract. However, a variety of external heating sources can affect the planet during its life. These can include heating the planetary surface through irradiation by its host star;  tidal heating induced by the star when the planet migrates, or any other potential heating sources resulting from other interactions and dynamical processes (e.g. collisions with other planets, \citealt{LinIda1997}, which can affect the early stages of planetary evolution and growth). Here, we consider the evolution of fully-formed planets after they had been excited to a high eccentricity, such that they experience strong tidal interactions with the host star. Besides the initial excitation to high eccentricity, defining our initial conditions, and the tidal interaction with the host star, we assume that no further interaction with other stellar or planetary bodies occur. Fig. \ref{fig:free contraction} demonstrates the fast contraction from the initial inflated radii to about $2 R_J$, in less than a $Myr$, such that a scattering prior to this stage is less probable, and even in such cases the binary would more likely be disrupted rather than rapidly migrate to produce a HJ/WJ (see discussion on flow in parameter space in paper-I). Therefore, we begin our models after a gas-giant have already finished the core accretion stages and any planetary scattering epoch, and reached the initial eccentricity for its migration. Generally, these processes are thought to have been finalized by the first few Myrs of evolution. As we describe in Subsec. \ref{sec:numerical}, we examine different initial radii at the time of coupling, such that the external energies are included both during the rapid contraction shown in Fig. \ref{fig:free contraction}, and after all initial models have been already converged and continue with the same cooling timescale. 

Hereafter we study the effects of two sources of external energy: tidal heating and irradiation flux from the host star, both of them taken into consideration self consistently together with the migration of the planet towards the host star, and the thermal cooling of the migrating planet. 

The distribution of the heat from the different sources inside the planet depends on the specific mechanism and the internal structure of the planet. Irradiation flux heats the surface of the planet and dissipates to deeper layers, but tidal heating may cause a deformation of the internal structure, and therefore can potentially heat deeper layers more efficiently. 
We define $r_{\rm ext}$ as the radial distance inside the planet in which most of the external energy source is deposited. 

The irradiation luminosity (averaged over an orbital period), is deposited in the photosphere of the planet( i.e $r_\text{ext} = r_\text{irr}$ = $R_p$ where $R_p$ is the radius of the planet), is given by: 
\begin{align} \label{eq:L_irr}
\bar L_{\rm irr} = \frac{1}{\mathcal{T}}\int_{\mathcal{T}} L_{\rm irr}(r(t))dt=\left(\frac{R_p}{a}\right)^2 \frac{L_\star}{\sqrt{1-e^2}}
\end{align}
where $\mathcal{T}$ is the orbital period and $r(t)$ is the distance between the planet and its host star.

The energy from tidal heating is given by the tidal model, which determines its internal distribution  (i.e $r_\text{ext} = r_\text{tides}$). We discuss the different heat distributions in Sec. \ref{subsec: equilibrium tide model} and \ref{subsec: dynamical tides} for the equilibrium and dynamical tides models. We explain our numerical method of the internal heat distribution in Sec. \ref{sec:numerical}.

We find that due to the planet's own radiation and cooling of the planet, the effect of deposition of irradiation and/or tidal heating on the dynamical evolution is mostly negligible when the energy is deposited in the planetary photosphere. In this case, most of the deposited energy is quickly irradiated away, and do not heat the planetary interior. Consequently, the planetary radius is not affected by the heating processes in this case nor affect the migration timescale. However, some processes, such as the Ohmic dissipation \citep{BatyginStevenson2010}, can provide a channel for heat conduction into internal regions. Deeper deposition at the inner layers could lead to much more significant effect, such that even $1\%$ of the external energy deposited at the center of the planet could induce larger radii than a $R_J$ even after $\rm Gyrs$ when the planet is already very close to its star (see \citealt{Bodenheimer2001, GuillotShowman2002, Komacek2020} and references therein). When using $r_\text{ext}=0$, to deposit the energy around the center of the planet, and multiply the right side of equation \ref{eq:extra_heat_distribution} by an efficiency parameter, we find that a very high energy deposition in the core can indeed give rise to planetary inflation, as can be seen in Fig. \ref{fig:dynamical1.5au0.98}, \ref{fig:compare_both_tides_1au0.97} and \ref{fig:dynamical_with_dep}, which in case of a strong inflation can lead to disruption. We further discussed here in Subsec. \ref{subsec:heat_tranfer}; see also the semi-analytic study in paper I.

\subsection{High Eccentricity Tidal Migration}
\label{sec: high eccentricity tidal migration}
Tidal migration occurs when strong tidal forces from the star act to exchange energy and angular momentum between the orbit and the planet, leading to the growth of tidal bulges and thus an orbital decrease.
Given the strong dependence of tides on the distance from the host star (Eq. \ref{eq:weak_tides}, \ref{eq:dynamical tides}), efficient tidal migration requires a close approach of the planet to the star. If the planet is born far from the star, as expected for gas-giants, a close approach can occur only if the planet resides on highly eccentric orbit, for which the pericenter approach is close to the star for tidal effects to become significant. Therefore, one can divide high eccentricity tidal migration into two separate stages: reducing the planet's angular momentum and reducing the planet's energy. In the first stage, the HJ/WJ progenitor which is likely formed on a relatively circular orbit is excited into a an eccentric orbit via planet-planet scattering\citep{RasioFord1996,Chatterjee2008,JuricTremaine2008}, as we discuss here, or through other channels for eccentricity excitation such as via Von-Ziepel-Lidov-Kozai (ZLK) mechanism and secular chaos \citep[e.g][]{VonZeipel1910,Lidov1962,Kozai1962,WuMurray2003,FabryckyTremaine2007,Naoz2011HJ,Petrovich2015a,Nagasawa2008,WuLithwick2011, HamersLithwichPeretsZwart2017,Wu2018}.  
In the second stage, energy extraction via tides leads to a migration and circularization of the planet's orbit. The energy extracted from the orbit during an orbital period is dissipated in the planet affecting its overall luminosity, which can affect the internal structure as a result. Assuming a complete transfer from the orbital energy to the planet, the injected/incoming luminosity can be described as follows:
\begin{align}\label{eq:L_tide}
L_{\rm tide} = -\frac{E}{a}\frac{da}{dt}
\end{align}
where $E$ is the orbital energy and $a$ is the semimajor axis. 

Modeling tides in giant planets is not trivial, and its strong dependence on the internal structure of the planet, turbulent viscosity processes dissipating energy and other physical aspects of the problem has induces some long-standing debates on the nature and specific properties of tidal dissipation. Here we adapt the widely used tidal model of weak/equilibrium tides \citep{Darwin1879,GoldreichSoter1966,Alexander1973,Hut1981}, but also consider more briefly the importance of dynamical tides \citep[e.g.][]{Zahn1977, Mardling1995a,Mardling1995b}. The latter could be especially important and more efficient during the early migration phases when the planet orbit is still highly eccentric, and in that sense, considering only weak-tides model is potentially conservative in term of the efficiency of eccentric migration\citep{Lai1997}.

Here we present a general approach, which can account for any tide model, and is demonstrated here using both equilibrium tides and dynamical tides. We note that other models such as chaotic-dynamical tides\citep{VickLai2018,VickLaiAnderson2019}, are likely to further shorten the migration timescales; these are to be left for future works. 

In the next sub-sections we explain how we model the migration of a planet due to equilibrium and dynamical tides, where we describe the equations of motion and the corresponding heat that should be transferred to the planet.

\subsubsection{Equilibrium tide model}\label{subsec: equilibrium tide model}
In this tidal model, the gravity from the star raise tides on the planet leading to formation of an equilibrium bulge on the planet, which is treated as external point mass along the calculation \citep{Hut1981}. Due to the timescale involved in raising the bulge, and the spin of the planet, the bulge position lags in respect to the position of the star, and the mutual interaction of the stellar gravity and the bulge torques the planet. 
When the lag time between the objects is much smaller than the spin or orbital period of the planet, one can invoke the weak tide approximation\citep{Darwin1879,GoldreichSoter1966,Alexander1973,Hut1981}.
Under the assumption of pseudo-synchronization (of the planetary spin and the orbit) and a conservation of the angular momentum, the orbital-averaged time evolution of the eccentricity and semimajor axis is given by \citep{Hut1981,HamersTremaine2017}
\begin{align}\label{eq:weak_tides}
\frac{da}{dt}=& -21 k_{\rm AM} n^2 \tau_p \frac{M_\star}{M_p}\left(\frac{R_p}{a}\right)^5ae^2 \frac{f(e)}{(1-e^2)^{15/2}}, \\
\frac{de}{dt} =& -\frac{21}{2}k_{\rm AM}n^2 \tau_p \frac{M_\star}{M_p} \left(\frac{R_p}{a}\right)^5 e\frac{f(e)}{(1-e^2)^{13/2}}
\end{align}
where $M_\star$ is the mass of the host star, $M_p, \ R_p, \ e, \ a, \ n$ and $\Omega_p$ are the mass, radius, orbital eccentricity, orbital semimajor axis mean motion and spin frequency of the Jupiter correspondingly; $\tau_p=0.66 \ \rm sec$ is the planetary tidal-lag time and $k_{\rm AM}=0.25$ is the planetary apsidal motion constant \citep{HamersTremaine2017}, and 
\begin{align}
f(e) := \frac{1 + \frac{45}{14} e^2+ 8 e^4 + \frac{685}{224} e^6 + \frac{255}{448} e^8 + \frac{25}{1792} e^{10}}{1 + 3 e^2 + \frac{3}{8} e^4}
\end{align}

Here we neglect the influence of the tides on the host star, as these are typically negligible in comparison with the tides on the planet. The energy associated with the tides according to equations \ref{eq:L_tide} and \ref{eq:weak_tides} scales as $R_p^5$, leading to a very strong dependence of the migration timescale on the planet's radius. Consequently, the migration timescales of initially inflated gas-giants should be shorter than the timescales of non-inflated gas-giants with a constant $R_J$ radius. We note that the contraction timescales are sufficiently long to maintain inflated gas-giants throughout a significant part of their dynamical evolution, such that the initial radius of a HJ/WJ will leave a signature on its expected final parameters, that could be also observed.

We consider the location of the tidal bulge, given by \citep{MurrayDermott1999}
\begin{align}
    h_\text{weak} = \frac{M_\star}{M_p} R_p \left(\frac{R_p}{a}\right)^3
\end{align}
implying a peak of the external heat from tides at $r_\text{tides} = R_p - h_\text{weak}$ from the center of the planet.

\subsubsection{Dynamical Tides}\label{subsec: dynamical tides}
At very large eccentricities, tidal energy mostly dissipates near periastron, raising a large tidal bulge on the primary (the giant planet in our case). Consequently, such tidal evolution cannot be parametrized by its average over the entire orbit, as done in the equilibrium tide model \citep{MoeKratter2018}. The energy associated with this tidal deformation might excite internal energy modes of the planet (mainly the fundamental f-mode), which might induce an enhanced response \citep{Mardling1995a,Mardling1995b,Lai1997,Ogilvie2014}, potentially leading to even more rapid circularization and migration of the planet. The eccentricity decay is accompanied by pseudo-synchronization with the angular frequency of the host star and the excitation of oscillations in the planet become less pronounces as the orbital eccentricity decreases. As a result, the energy dissipation by the various modes is gradually suppressed, until a transition to the regime in which equilibrium tides are more dominant\citep{Mardling1995b}. The quadrupole order of the energy dissipation can be written as follows: \citep{PressTeukolsky1977,MoeKratter2018}
\begin{align}\label{eq:E_dyn}
\Delta E = f_{\rm dyn}\frac{M_\star+M_p}{M_p}\frac{GM_\star^2}{R_p}\left[\frac{a(1-e)}{R_p}\right]^{-9}
\end{align}
with $f_{\rm dyn}=0.1$, as \cite{MoeKratter2018} (following the calculation of \citealt{McMillan1986}) found in good agreement with observations of pre-MS, which are approximated by the same polytropic index $n=3/2$  assumed for gas giants. We note that these are estimates containing a large uncertainty, and the exact displacement of this energy is still not well understood. We test the implication of such choice in the result section. 

At very high eccentricities, the orbital angular momentum can be very low, and one might not neglect the spin angular momentum as done in the above prescription of the equilibrium tides. Therefore, during the migration stage dominated by dynamical tides, one can assume a conservation of the pericenter instead of the angular momentum \citep{MoeKratter2018}.
Combining this prescription with the equations of the orbital energy and angular momentum, and assuming a constant pericenter, leads to the following equations of the orbital semimajor axis and eccentricity along the migration\citep{MoeKratter2018}:
\begin{align}\label{eq:dynamical tides}
\frac{da}{dt} = -\frac{a}{\mathcal{T}} \frac{\Delta E}{E}, \\ 
\frac{de}{dt}= \frac{1-e}{a}\frac{da}{dt}
\end{align}
Where $\mathcal{T}=2 \pi \sqrt{\frac{a^3}{G\left(M_\star+m_p\right)}}$ is the orbital period of the planet around its host star.

\begin{figure}
\centering
\includegraphics[width=\linewidth]{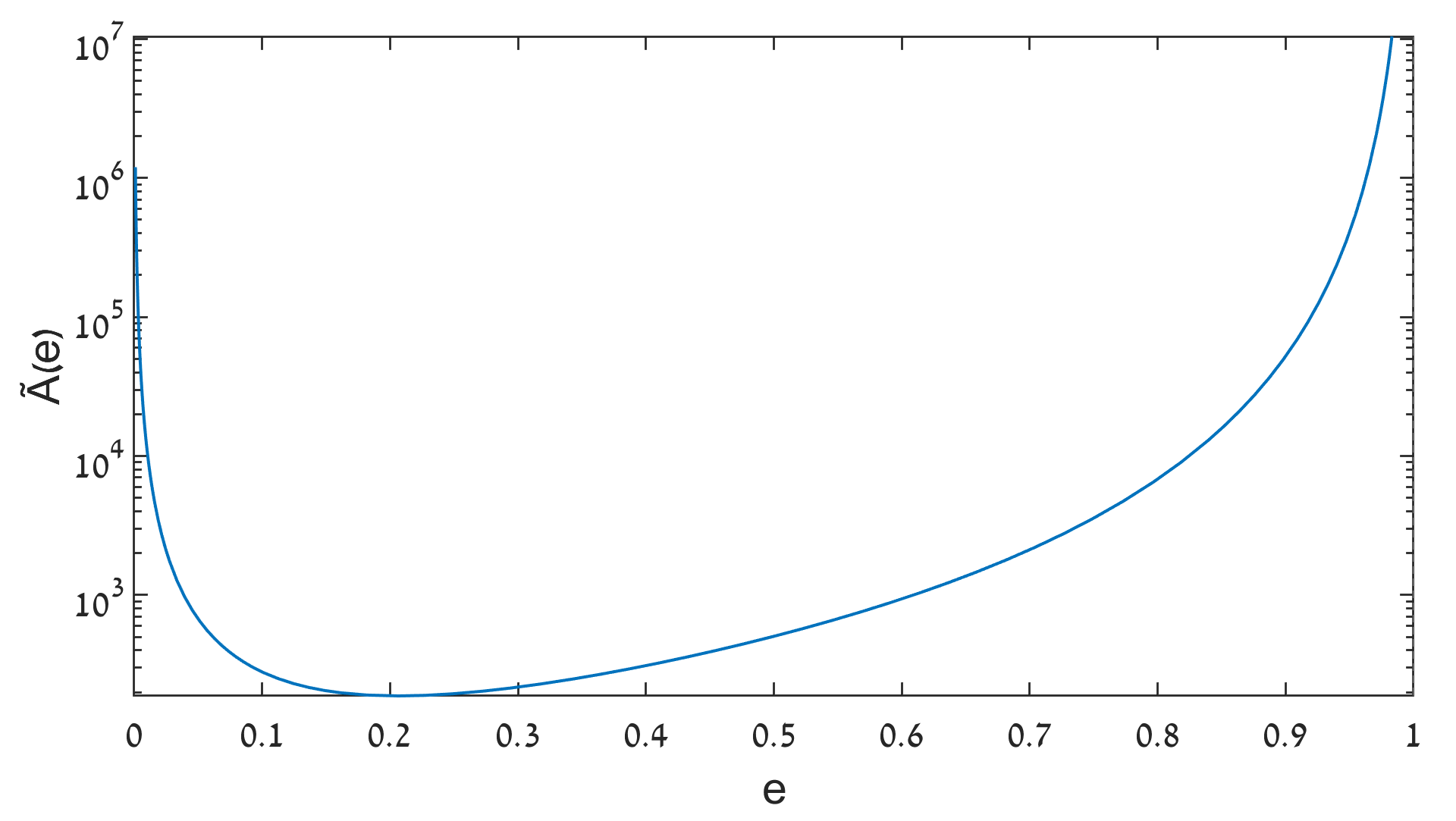}
\caption{$\Tilde{A}(e)$ from Eq. \ref{eq:new_beta} showing the dependence of the dynamical-to-weak tide migration strengths ratio ($\beta$) on the eccentricity. }
\label{fig:A_Tilde}
\end{figure}

While dynamical tides dominate for large eccentricities, weak tides will be a more physical description for low ones \citep{Mardling1995b}.
The ratio of the migration rate due to dynamical tides to the migration rate due to weak tides is given by
\begin{align}
\beta(R_p,a,e) \equiv \frac{da/dt|_{\rm dyn}}{da/dt|_{weak}}= \frac{2f_{\rm dyn}R_p^3 A(e)}{21 GM_p k_{AM}\tau_p \mathcal{T}}
\end{align}
where 
\begin{align}
A(e) \equiv \frac{(1-e^2)^{15/2}}{(1-e)^9 e^2 f(e)}
\end{align}
Since the pericenter is assumed to remain constant during the migration with dynamical tides, one can write:
\begin{align}
\beta(R_p,a,e) = \Tilde{A}(e) \cdot R_p^3 \cdot B(m_p,...)
\label{eq:new_beta}
\end{align}
where $\Tilde{A}(e) \equiv A(e) \cdot \left(1-e\right)^{-3/2}$.
The transition between the dynamical and weak tides occurs roughly at $\beta \sim 1$, and we set a lower artificial cutoff at $e=0.2$, at approximately the point where $\Tilde{A}(e)$ gets its minimal value (see Fig. \ref{fig:A_Tilde}). In this way, we avoid the divergence of dynamical tides at $e=0$, and the transition occurs at $\max\{0.2,e|_{\beta=1}\}$. We note that considering migration due to dynamical tides with a very small pericenter such that $B$ in Eq. \ref{eq:new_beta} is greater than the maximum value of $\frac{1}{\Tilde{A}(e)\cdot R_p^3}$, leads to $\beta>1$ for the entire migration until $e=e_\text{trans}$.  We further discuss the transition point between the different tide models in Subsec. \ref{subsec:diff_tides}.

We note that at a sufficiently large eccentricity and low pericenter, the oscillatory modes inside the planet due to tides could grow chaotically \citep{IvanonPapaloizou2004,IvanovaPapaloizou2007,VickLai2018,Wu2018}, and can potentially increase the energy exchange and hence lead to a faster migration and circularization. However, modeling such scenario will be left for a future work. 

In our modeling, the energy from the dynamical tides is deposited into the planet's photosphere, i.e - $r_{tides}=R_p$. However, more accurate future models might include different internal distribution, as the deposition heat from dynamical tides inside a planet is not yet understood \citep{Sun2018}, and is beyond the scope of this work. In Subsec. \ref{subsec:heat_tranfer}, we briefly discuss the effect of a deeper deposition of the tides, which may arise from an efficient Ohmic dissipation.

We compare the evolution of inflated eccentric migrating planets due to two different tide models - weak and dynamical tides in Sec. \ref{subsec:diff_tides}, where one can notice the stronger effect of the dynamical tides, not only on the migration process, but on the planet's structure as well. In the same sub-section, one can find a further discussion on the differences between the different tide models, in addition to the influence of different parameter choices for the dynamical-tides model.

\subsection{Numerical coupling of the thermal-dynamical evolution}

\label{sec:numerical}
In our numerical approach, we couple the orbital averaged equations, derived from the tidal migration model, with a numerical modeling of the internal evolution of the planet, affected by the deposition of heat and the cooling due to  its own irradiation. We use the \texttt{AMUSE} framework \citep{Portegies2009AMUSE} (version 13.2.1 including self contributions that should be available in future versions) to combine between the different codes. The internal evolution of our planets is modeled with the stellar evolution code \texttt{MESA} \citep{PaxtonMESA2011, PaxtonMESA2013} version 2208, which is a one-dimensional code, that solves the stellar equations \citep{Kippenhahn2012} assuming a hydrostatic equilibrium, and a spherical symmetry. We use the OPAL/SCVH equations of state \citep{Rogers2002OPAL} and opacity tables corresponding to existence of molecules in low temperature at the outer layers of the planets \citep{Grevesse1998GS98,Freedman2008}. In order to simulate such planets, our code can combine any given opacity tables by specifying the transition temperatures. In this way, one can consider available dust opacities as well as any future opacity tables that are relevant for planets and were not available in the original version of \texttt{MESA} within \texttt{AMUSE}. 
\begin{figure}
\centering
\includegraphics[width=\linewidth]{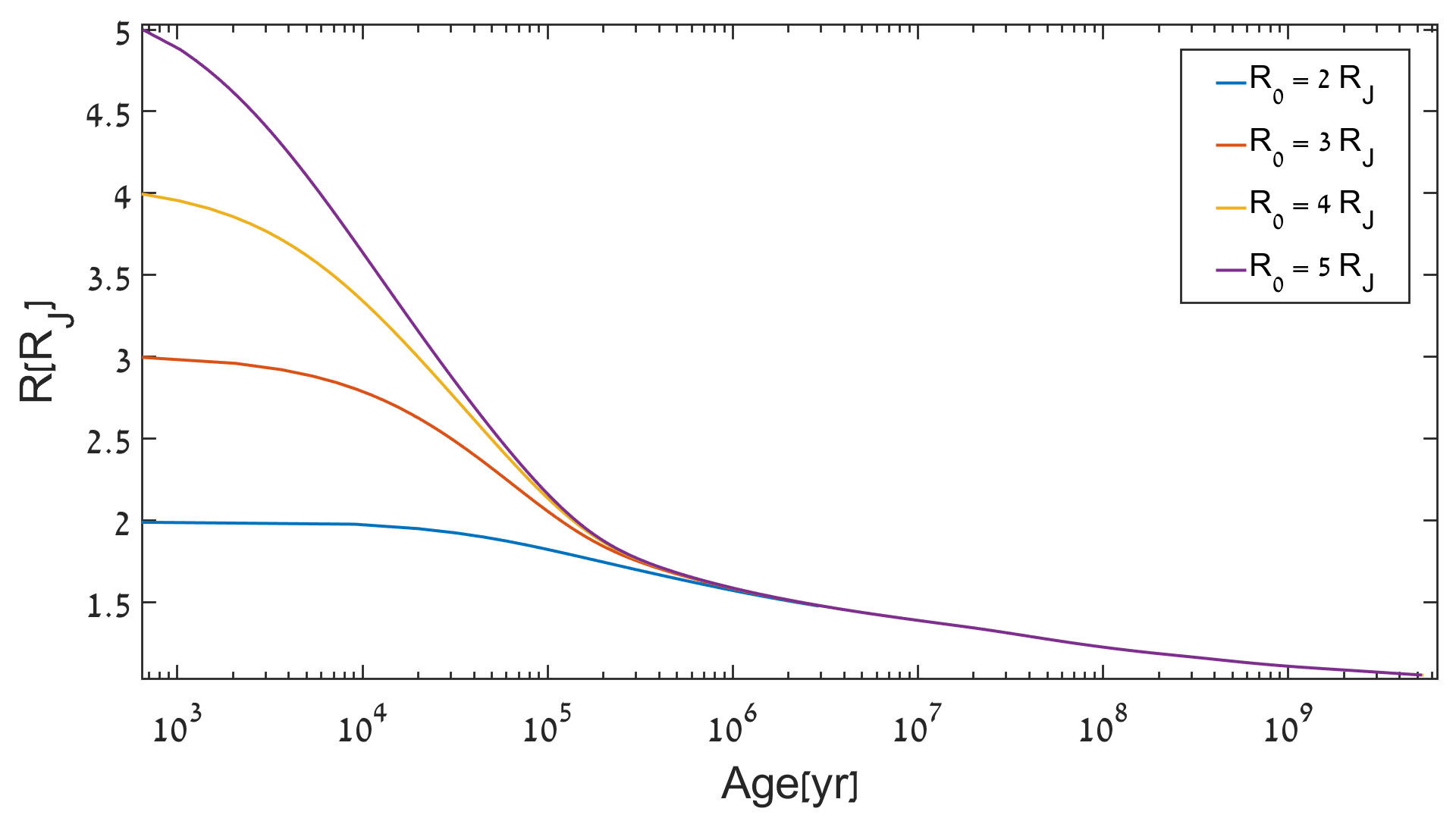}
\caption{Contraction of a gas giant, modeled with \texttt{MESA} version 15140 \citep{PaxtonMESA2013,PaxtonMESA2018,PaxtonMESA2019}, with different initial radii, without deposition of external heat. 
}
\label{fig:free contraction}
\end{figure}
We begin by creating the initial planet model as a Pre-main-sequence low mass star, which has no nuclear burning and hence mimics the evolution of a planet. The planet model then contracts according to the equilibrium between its own gravity and thermal cooling, as can be seen in Fig. \ref{fig:free contraction}). We thus evolve this model in isolation until reaches the initial radius at which the migration process is assumed to begin, and from this point we couple its further internal evolution (i.e thermal cooling) to the dynamical evolution. When simulating the coupled migration process, after each orbital evolution step, we calculate the corresponding external heat source (extra heat as termed in \texttt{MESA}\citep{PaxtonMESA2013}), $L_\text{ext}$, such that the energy equation of the planet at each radial distance from its core becomes:
\begin{align}
    \frac{dL}{dm} = -T\frac{ds}{dt} + \frac{dL_\text{ext}}{dm} 
\end{align}
where $T$ is the temperature of the model and $s$ is the specific entropy, and $L_\text{ext}$ is calculated according to the orbital parameters, through the dependency of tides and irradiation, described in the previous sections. In order to include the external heat term via \texttt{AMUSE}, we updated the current interface to support the inclusion of any external heat distribution during the evolution with \texttt{MESA}.

We consider a heating source which deposits its energy at some typical region inside the planet, at a distance of $r_\text{ext}$ from the planet's center. For lack of a known distribution of the heat, we adapt a Gaussian heat distribution similar to \cite{Spiegel2013, KomacekYoudin2017}:

\begin{align}\label{eq:extra_heat_distribution}
    \frac{dL_\text{ext}}{dm} = \frac{L_\text{ext}}{\sqrt{2\pi \sigma_\text{ext}^2}}\exp{\left[-\frac{1}{2}\left(\frac{r-r_\text{ext}}{\sigma_\text{ext}}\right)^2\right]}\frac{dr}{dm}
\end{align}
 where $dr/dm=(4\pi \rho r^2)^{-1}$  and $\sigma_\text{ext} = 0.5H_{p,\text{ext}}$ is half the scale height, computed at $r_\text{ext}$ according to \citep{PaxtonMESA2011}:
 \begin{align}
     H_p\left(r\right) = \begin{cases}\min\left\{  \frac{P}{\rho g}, \sqrt{\frac{P}{G\rho^2}} \right\}  & ,r\ne 0 \\
     R_p & ,r=0
     \end{cases}
 \end{align}
 where $P$ and $\rho$ are the pressure and density distributions of the planet. Since this distribution depends on the pressure and density profiles, a deposition at a higher location in the atmosphere, at low pressure regions, is likely to have only a little influence on the interior of the planet, whereas deposition of energy directly into the core significantly affects the evolution of the planetary structure. Integrating over eq. \ref{eq:extra_heat_distribution} gives approximately the same amount of heat as $L_{\rm ext}$. We note that the sum of this discrete numerical distribution over all shells might be different (lower) than the total heat calculated from eq. \ref{eq:L_irr} and eq. \ref{eq:L_tide}, when using a Gaussian distribution around some radial distance inside a spherical model with a discrete mass distribution.  The change we find is by a factor of 2 at most, i.e. this can effectively be translated to a lower efficiency of the heat conductance to the central parts. Other distributions can be chosen and easily used by changing the heat distribution.
 
The time of each step in our simulation is chosen to be much shorter than the typical timescales for the orbital/thermal changes. 
At each step, we use the current properties of the planet, and evolve the orbital parameters according to the specific tides model (i.e Eq. \ref{eq:weak_tides} or \ref{eq:dynamical tides}). 

The amount of deposited energy by tides is derived from the tides model, and is deposited in the planet, see eq. \ref{eq:L_tide}. 
We assume the energy is deposited at a typical radius of the planet, as discussed above, and smooth it as a Gaussian distribution of the corresponding heat\citep{KomacekYoudin2017} $L_\text{ext} = L_\text{tides}$, with a pick at $r_\text{ext} = r_\text{tides}$, inside the \texttt{MESA} model, as described in Eq. \ref{eq:extra_heat_distribution}, where the exact $r_\text{tides}$ depends on the tides model, as described in sec. \ref{sec: high eccentricity tidal migration}.
 The irradiation flux from the host star, which changes due to the orbital evolution throughout the migration process is described in Eq. \ref{eq:L_irr}. The corresponding heat is distributed in the photosphere of the planet , using $r_\text{ext}=R_p$ in Eq. \ref{eq:extra_heat_distribution}. 
 
  After injecting both tidal and radiation energies , i.e- $L_\text{ext}=L_\text{tides}+L_\text{irr}$, we evolve the planet model with \texttt{MESA} for the same duration as was done for the orbital evolution. Our simulations terminate when one of the following conditions is fulfilled: (1) the planet has passed its Roche limit, defined as $r_t = \eta R_p \left(\frac{M_\star}{M_p}\right)^{1/3}$ \citep{Guillochon2011}, and can not survive in its current condition, (2) the evolution time has passed the Hubble time, (3) the planet has cooled and contracted to the levels such that currently used finite opacity and EOS tables, as well as other parameters in the used version of \texttt{MESA} are no longer adequate. Satisfying the last criterion (3) means that the contraction timescale of the current model (including external heating) is much shorter than the migration timescale. During the migration this termination condition was achieved only in some of our simulations which produced WJs. 
  In such cases, the numerical model can not be compared with the semi-analytic model throughout the evolution. Modeling of these regimes can, however, be followed in the the semi-analytical model (see paper I). 
  
Our numerical model has been developed such that one can choose different orbital evolution models and different internal evolution codes. Using the current \texttt{MESA} module, one can follow the evolution of an externally built \texttt{MESA} model, and include any external heat sources distributed around any desired location according to \ref{eq:extra_heat_distribution}.

\section{Results}
\label{sec:Results}
In the next section we present the results of our numerical simulations of the inflated eccentric migration of gas-giants. We simulated different candidate models producing HJs\&WJs, where we tested the effect of the different heat sources on their migration under the  different tide models.  

In paper I we present a more simple, semi-analytical model, that uses the same equations of motion for the orbital change, but is coupled with equations approximating the thermal/radius change, instead of coupling to the much more detailed, yet computationally expensive numerical model of the internal evolution as done here. We compare the results of the semi-analytical model to those of our numerical model described here and find excellent agreement (see also paper I). 

\subsection{Hot \& Warm Jupiter candidates with different initial properties}
The thermal-dynamical evolution is affected by the different properties of the migrating planet; its mass, radius and the internal distribution of heat from external energy sources.

\begin{figure*}
\centering
\includegraphics[width=0.87\linewidth]{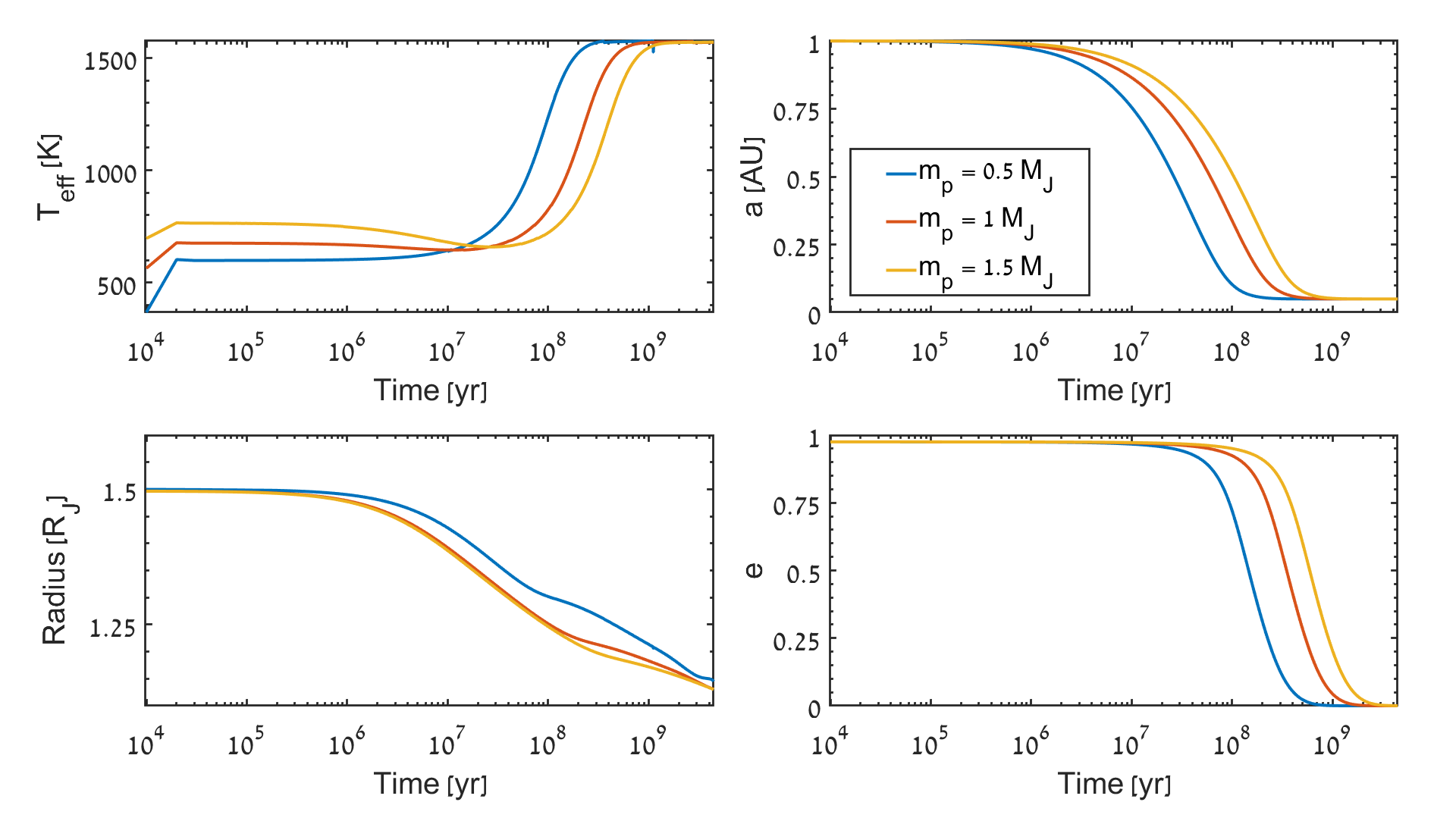}
\caption{
The numerical models' results of the thermal and orbital evolution of a HJ progenitor with different initial mass, migrating due to weak tides, including irradiation and tidal heating. The initial radius is $1.5 \ \rm{R_J}$, the initial semimajor axis is $1 \ \rm{AU}$ and the initial eccentricity is $0.975$.
 }
\label{fig:masses}
\end{figure*}

\begin{figure*}
\centering
\includegraphics[width=0.87\linewidth]{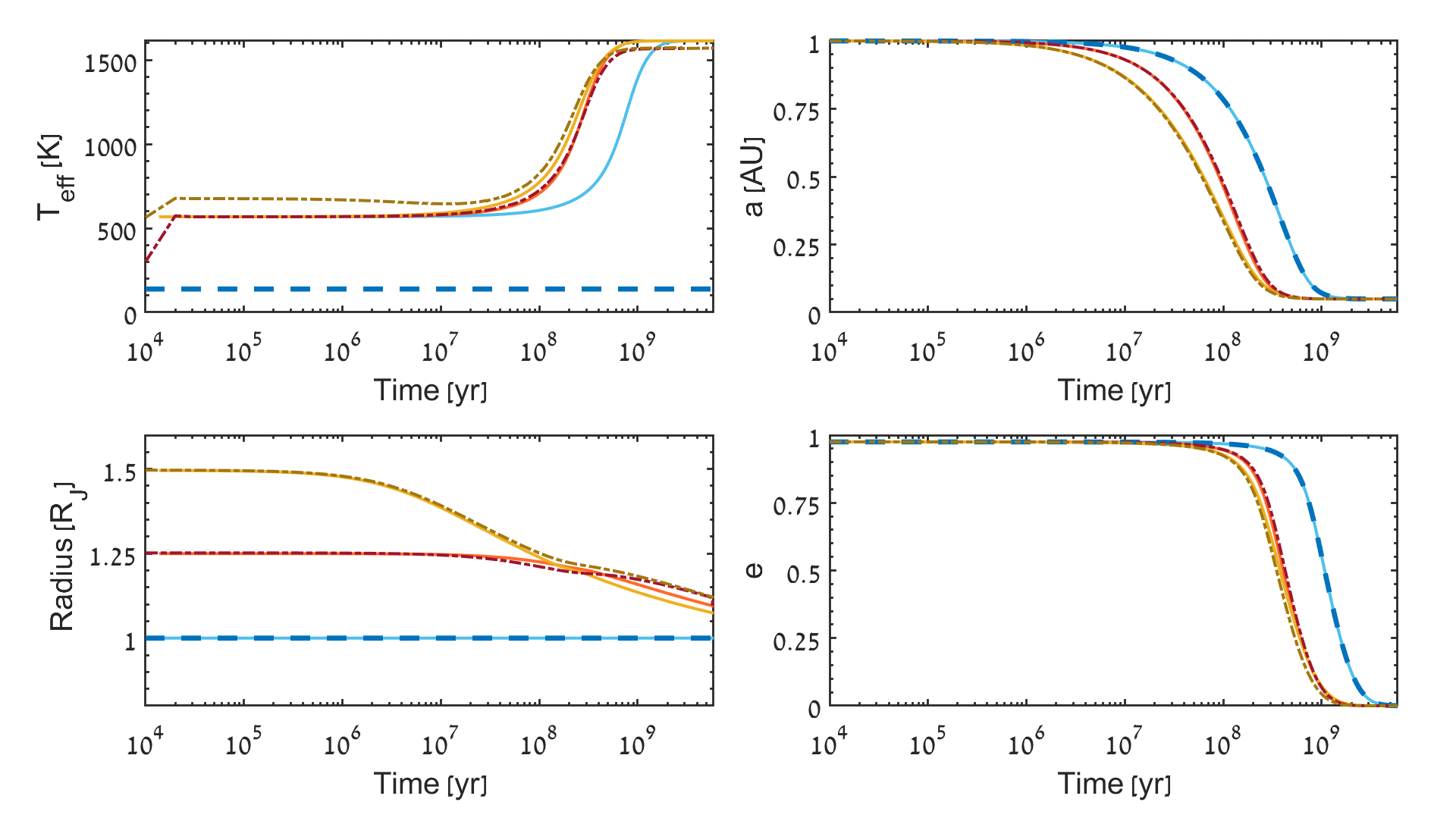}
\caption{
Same as Fig. \ref{fig:masses}, but all with same mass $1 M_J$ and different initial radii. The blue line corresponds to the the migration of a gas-giant assumed to have a constant $1 R_J$ throughout its evolution, without consideration of any thermal evolution, as typically done in eccentric migration models. The solid lines correspond to the results from the semi-analytical model and the dashed lines to results from the numerical model.  In light-blue is the semi-analytical simulation of a gas-giant with an initial $1 R_J$ radius, but now considering its thermal evolution.
}
\label{fig:weak1au0.975}
\end{figure*}

Observations show that the majority of the giant-planets population have mass in the range of $0.1-10$ M$_J$ \citep{Butler2006}.
In Fig. \ref{fig:masses} we present a comparison between the evolution of planets of different masses according to the equilibrium tides model, showing that planets with lower masses migrate faster. The slower contraction of the radii, shows inflated eccentric migration to have an even larger impact in these cases, when considering the formation of low-mass HJs\&WJs. This can be explained by the opposite dependence of the weak tides EOS on the planetary mass (Eq. \ref{eq:weak_tides}). In this case the resulting giants have migrated to become HJs with a final orbital period of $\sim 4$ days.

The strong dependence of tides on the radius of the migrating object implies a faster migration for larger initial radii. In Fig. \ref{fig:weak1au0.975} we compare the dynamical and thermal evolution of gas planets, considering weak tides, which are initialized with different radii. We notice that an initially more inflated planet can migrate an order of magnitude faster or even more, than a planet with a constant $1R_J$.
We discuss the possible implications of the different assumptions regarding the initial radii on the HJs\&WJs population in paper I.

\subsection{Different tide models}
\label{subsec:diff_tides}
\begin{figure*}
\centering
\includegraphics[width=0.87\linewidth]{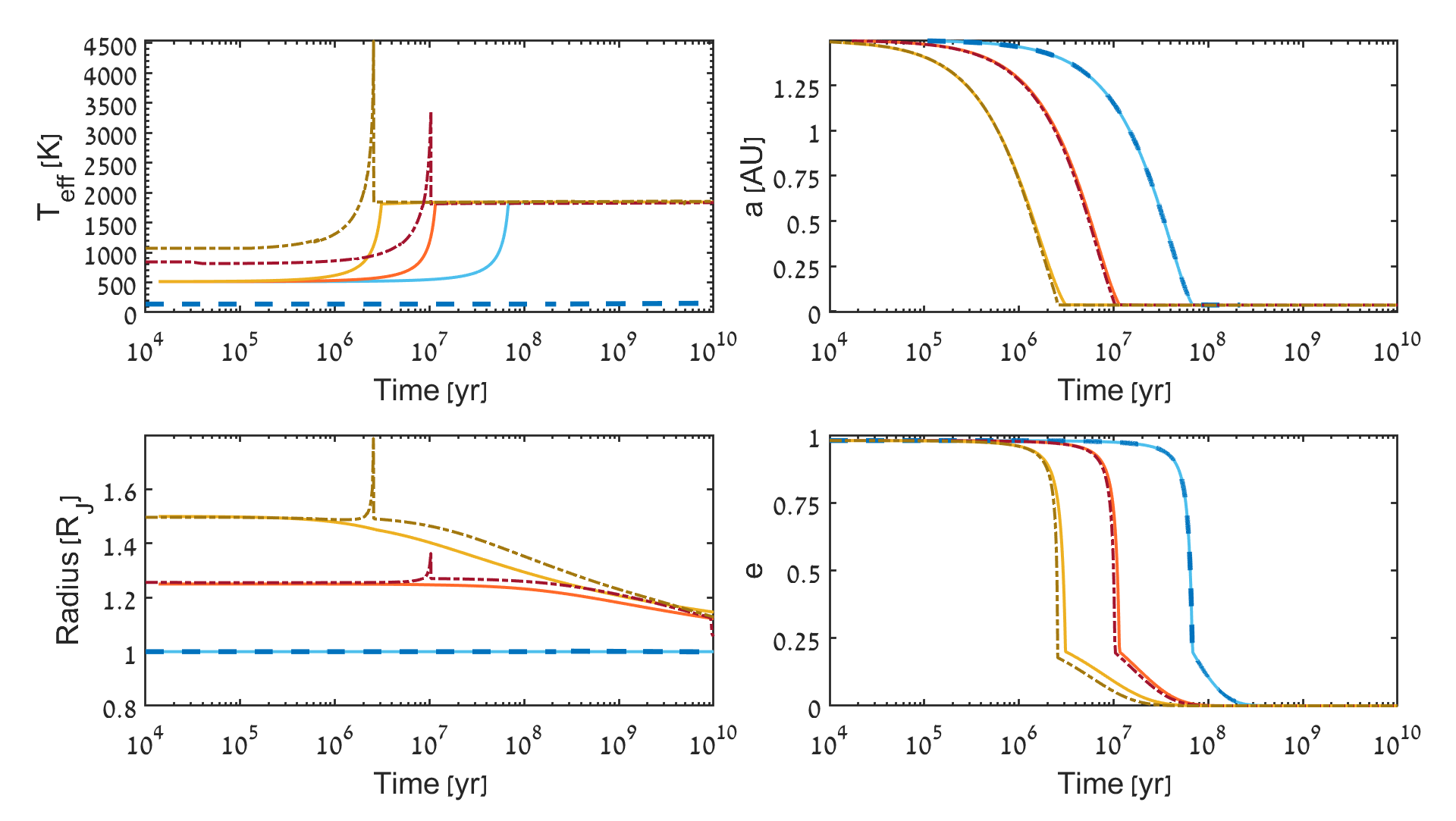}
\caption{
The thermal and orbital evolution of a HJ-progenitor migrating through dynamical tides. The initial semimajor axis is $a_0 = 1.5 \ \rm{AU}$,  and the initial eccentricity is $e_0 = 0.98$. Blue lines show the eccentric migration of a constant $1 R_J$ planet, without including its internal evolution; orange and yellow correspond to models with $R_0 = 1.25 R_J$ and $R_0 = 1.5 R_J$, in which we include the thermal evolution affected by both irradiation from the star and tidal heating. Dashed lines are the results of our numerical model compared with the straight lines derived with the semi-analytical approach presented in paper I, where light blue corresponds to the  semi-analytical model of a $1 R_J$ planet, when including thermal evolution. 
}
\label{fig:dynamical1.5au0.98}
\end{figure*}

The equations of motion of a migrating planet due to equilibrium tides are derived with the assumption of a small eccentricity. As was described in Sec. \ref{subsec: dynamical tides}, high eccentricities are likely to excite additional modes inside the planet that can lead to even larger effect on the migration process. In Fig. \ref{fig:dynamical1.5au0.98} we show the migration of gas-giant planets with different initial radii, but now affected by dynamical tides, compared with the migration of a constant $1 R_{\rm J}$ radius gas-giant. The migration of the initially inflated planets is indeed shorter by more than an order of magnitude compared with the migration of the constant radius planet.

\begin{figure*}
\centering
\includegraphics[width=0.87\linewidth]{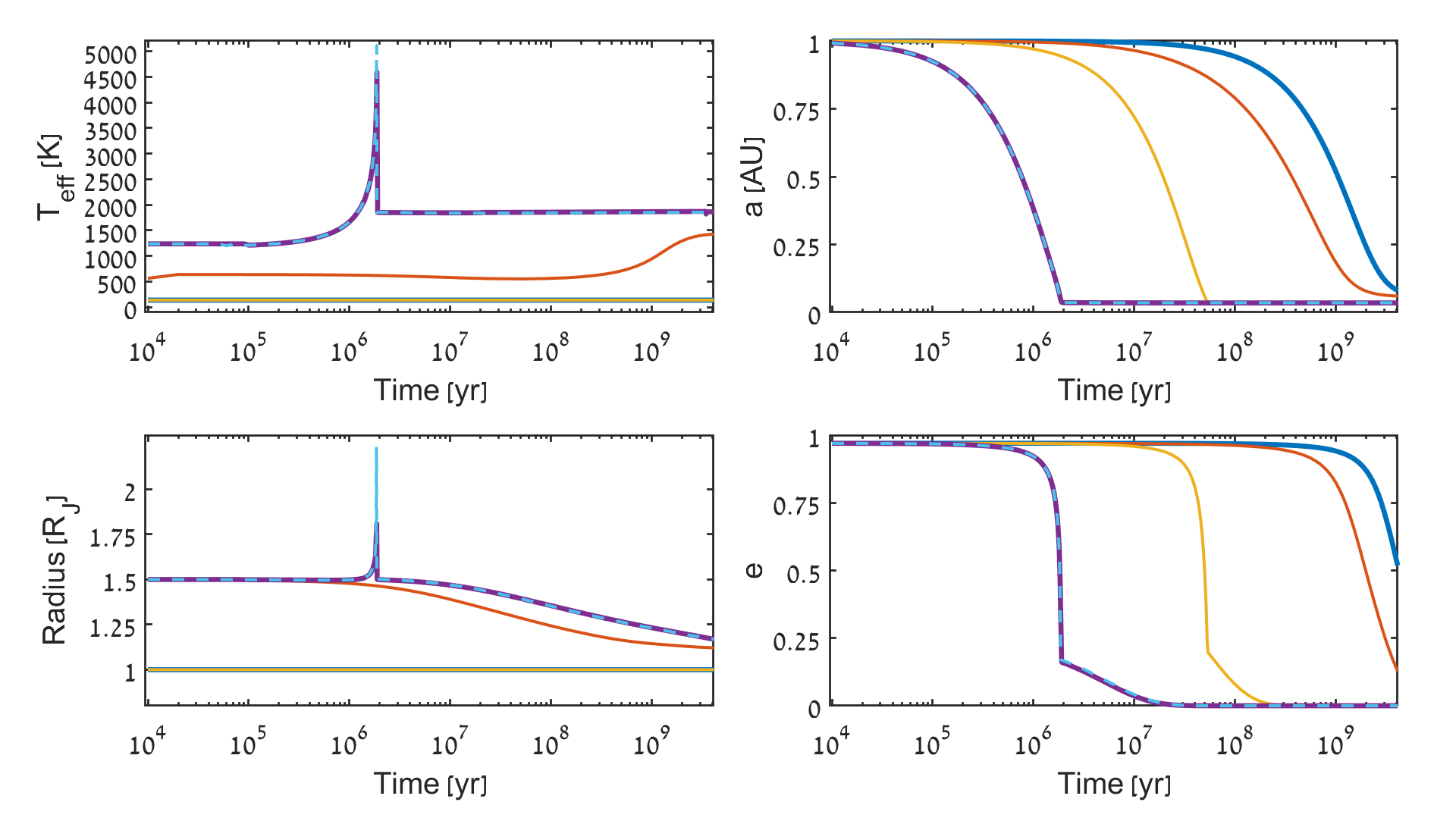}
\caption{Different models of migrating planets with $1M_J$ and initial orbital parameters of $a_0=1 \text{AU}$, $e_0=0.97$. Both models with initial $R_0= 1R_J$ (blue and yellow) are evolved only dynamically, with a constant radius, other models are evolved according to the different tide models, when including tidal heating and irradiation from the host star during their migration. Purple and yellow lines correspond to models evolved with dynamical tides, orange and blue correspond to equilibrium tide models. Light blue dashed line also evolves under the dynamical tides, but when the tidal heating is injected at a depth equals the size of the bulge for both tides models.}
\label{fig:compare_both_tides_1au0.97}
\end{figure*}

In Fig. \ref{fig:compare_both_tides_1au0.97} we compare the evolution of migrating planets when considering the two different tidal models, where one can notice the larger effect of the dynamical tides, with a greater dependence on the planetary radius.

\begin{figure}
\centering
\includegraphics[width=0.95\linewidth]{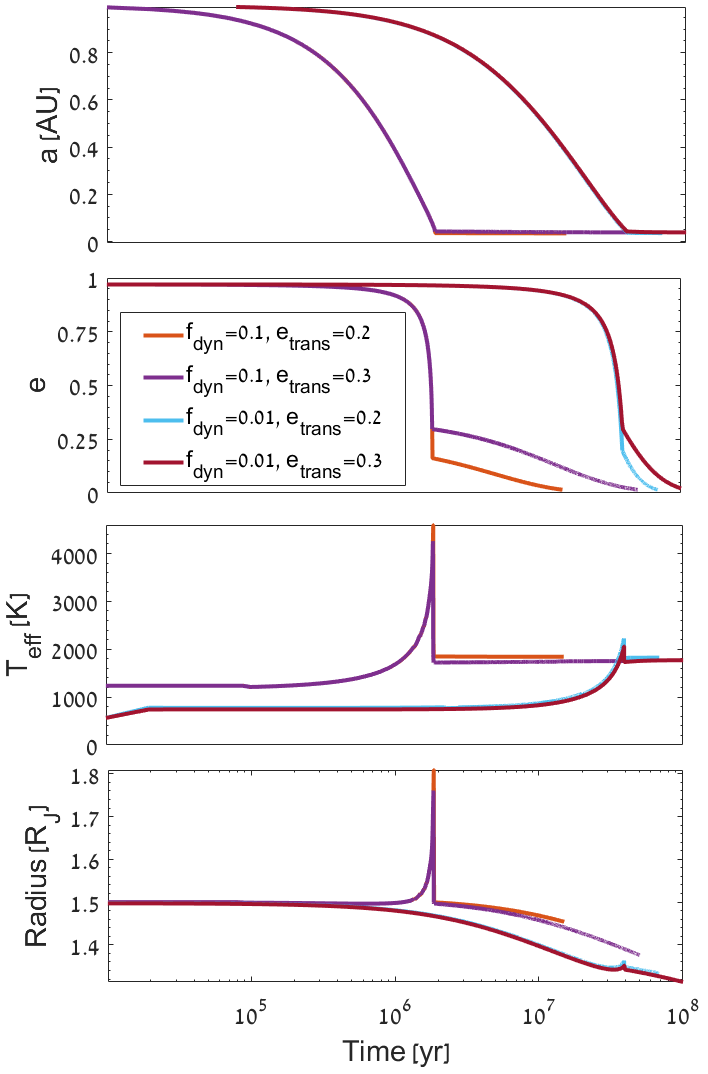}
\caption{Migrating planets with $1M_J$ and initial orbital parameters of $a_0=1 \text{AU}$, $e_0=0.97$, $R_0=1.5R_J$, affected by dynamical tides and irradiation from a sun-like host star.}
\label{fig:dynamical_1au0.97_diff_f_etrans}
\end{figure}

As described in Subsec. \ref{subsec: dynamical tides}, the dynamical tides model still has many uncertainties, among them are the value of $f_\text{dyn}$ in eq. \ref{eq:dynamical tides} \citep{McMillan1986,Mardling1995b,Lai1997} and the transition point to equilibrium tides \citep{MoeKratter2018,GRI22}. When using a value of $f_\text{dyn}=0.1$, and if $\beta > 1$ all the way to $e_\text{trans}$, the energy associated with the dynamical tides increases rapidly as the semimajor axis decreases. In this case,  even when the heat is deposited only at the planet's photosphere, sufficient heat is transported to the central part as to give rise to a radial expansion (re-inflation) of the gas-giant  (see Figs. \ref{fig:dynamical1.5au0.98}, \ref{fig:compare_both_tides_1au0.97} and \ref{fig:dynamical_1au0.97_diff_f_etrans}). This can be seen in Fig. \ref{fig:A_Tilde}, showing the dependence of $\beta$ on the eccentricity, which has a minimum at $~0.2$, and goes to infinity for $e\rightarrow 0$ and $e \rightarrow 1$. Since equations \ref{eq:E_dyn}, \ref{eq:dynamical tides} are no longer valid at $e=0$, one must use another condition to cease the dynamical tides prior to this point, in correspondence with Fig, \ref{fig:A_Tilde}, we choose a lower limit of the eccentricity with dynamical tides between $0.2$ and $0.3$. However, as there must be a smooth transition into equilibrium tides prior to circularization, the large jump in the tidal energy during the transition is probably not physical. In our model, we consider the heat associated with the equilibrium tides to be deposited at a depth of the bulge height, while the heat from the dynamical tides, which might be larger, is deposited around the surface, such that the depth of energy deposition changes after the transition, to a slightly deeper layer and a lower luminosity for the weak tides, as can be seen in Fig. \ref{fig:comb_L}. In Fig. \ref{fig:compare_both_tides_1au0.97} we also examine the evolution when $r_\text{ext}$ always equals the  difference between $R_p$ and the height of the bulge, also during the influence of the dynamical tides (light blue dashed line), and since the bulge is much smaller than the size of the planet, and therefore very close to the surface, the difference, compared with the evolution when dynamical tides are deposited around the surface (purple line), is relatively negligible. The same behaviour can be seen in Fig. \ref{fig:dynamical_with_dep}, when the external heat is deposited around the core for the entire evolution, which does not affect the dynamical evolution. In addition, as can be seen in Fig. \ref{fig:dynamical_1au0.97_diff_f_etrans}, lower value of $f_\text{dyn}$ leads to a smoother transition, with a lower impact of the dynamical tides on the migration. Fig. \ref{fig:dynamical_1au0.97_diff_f_etrans} shows only a minor difference in the eccentricity evolution between $e_\text{trans}=0.2$ and $e_\text{trans}=0.3$, but a significant difference in the migrations with the different efficiency parameters ($f_{dyn}$). We compare the effect of the efficiency parameter on the overall formation rate of the HJs\&WJs population in paper I, showing an increase in both populations when using larger values of $f_{dyn}$. We note that very large dynamical tides can lead to the planet disruption, even when still not reaching the tidal radius.

\subsection{The effect of different energy sources}
\begin{figure*}
\centering
\includegraphics[width=0.87\linewidth]{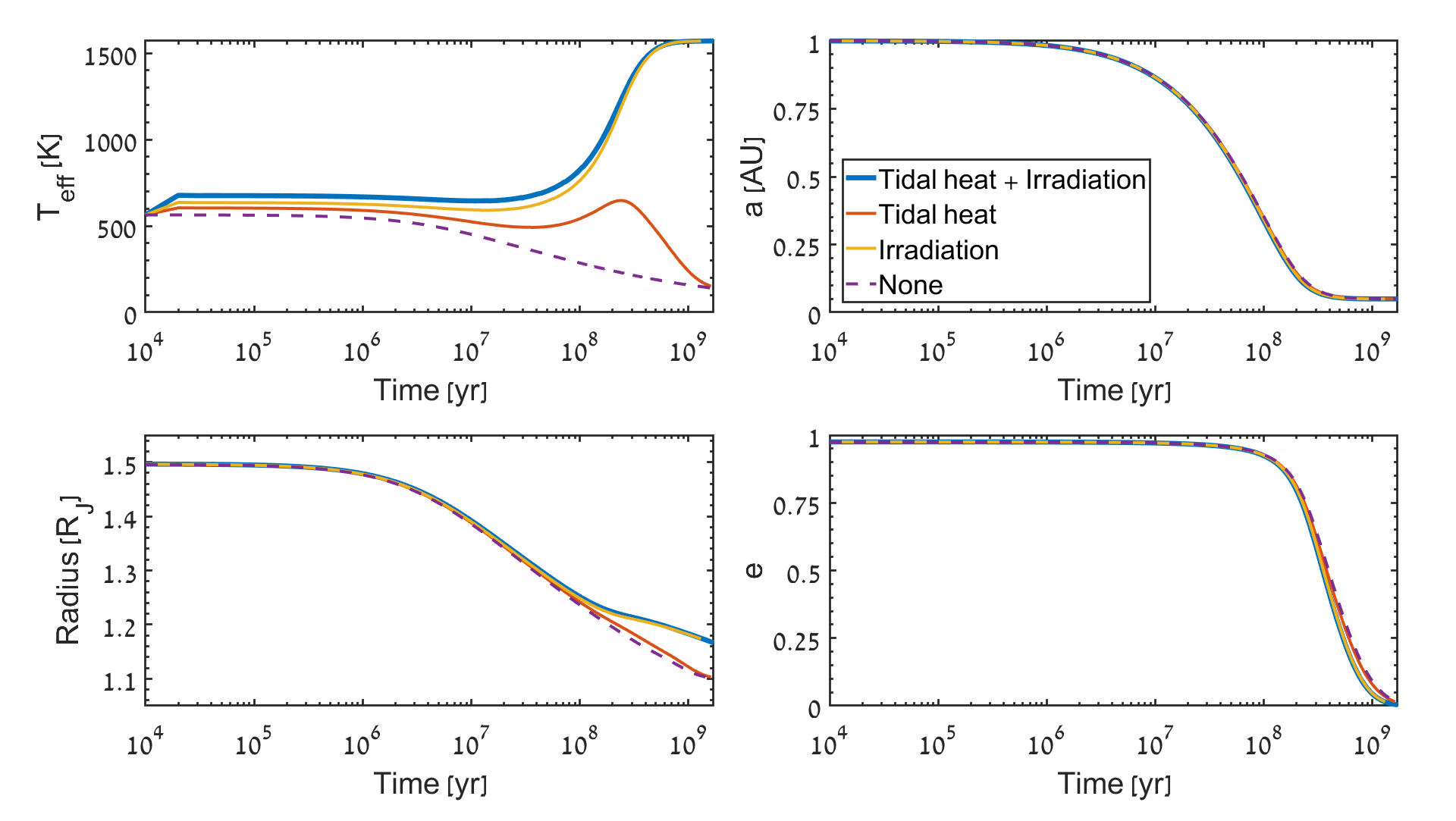}
\caption{Comparison of the thermal and orbital evolution of a HJ candidate migrating through weak tides. We consider the following initial conditions: initial radius of $1.5\rm{RJ}$; initial semimajor axis of $1 \ \rm{AU}$, and initial eccentricity of $0.975$ (partially presented in Fig. \ref{fig:weak1au0.975}), and we consider different external energy sources. Red lines corresponds to evolution without any external energies.  Orange line corresponds to evolution with (weak) tidal-heating; blue line to evolution with irradiation from the star, and the green line model includes both irradiation from the star and (weak) tidal heating. All the lines were produced with the numerical approach.
 }
\label{fig:weak_extras}
\end{figure*}

\begin{figure*}
\centering
\includegraphics[width=0.87\linewidth]{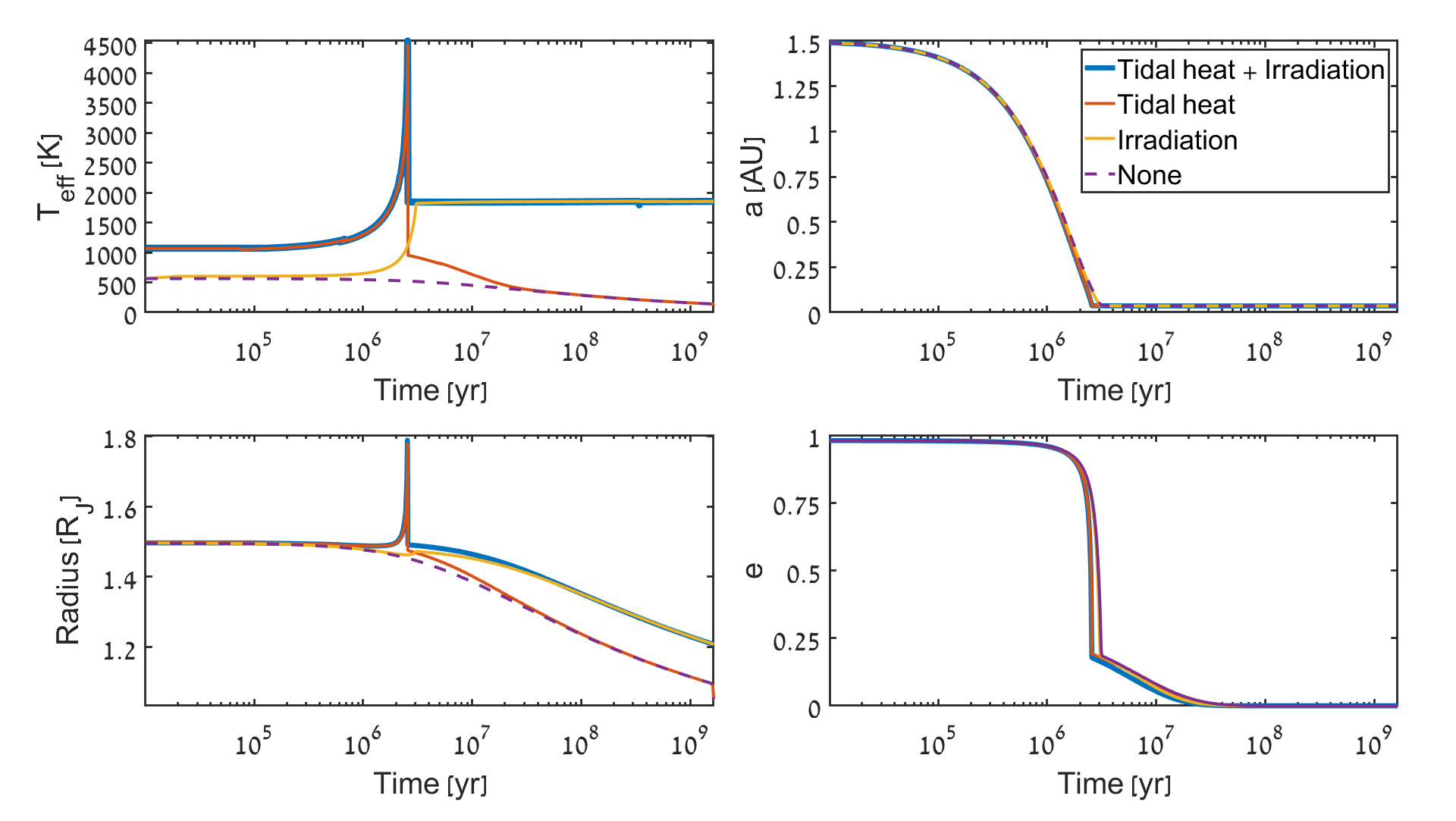}
\caption{Comparison of the thermal and orbital evolution of a HJ candidate due to dynamical tides.
We consider the following initial conditions:
initial radius of $1.5\rm{RJ}$; initial semimajor axis of $1.5 \ \rm{AU}$, and initial eccentricity of $0.98$ (partially presented in Fig. \ref{fig:dynamical1.5au0.98}). Different external energy sources are considered. The red line corresponds to evolution without any external energies; the orange one includes (dynamical) tidal heating, the blue line includes irradiation from the star and the green line includes both irradiation from the star and (dynamical) tidal heating. All the lines were produced with the numerical approach.}
\label{fig:dynamical_extras}
\end{figure*}

\begin{figure}
\centering
\includegraphics[width=\linewidth]{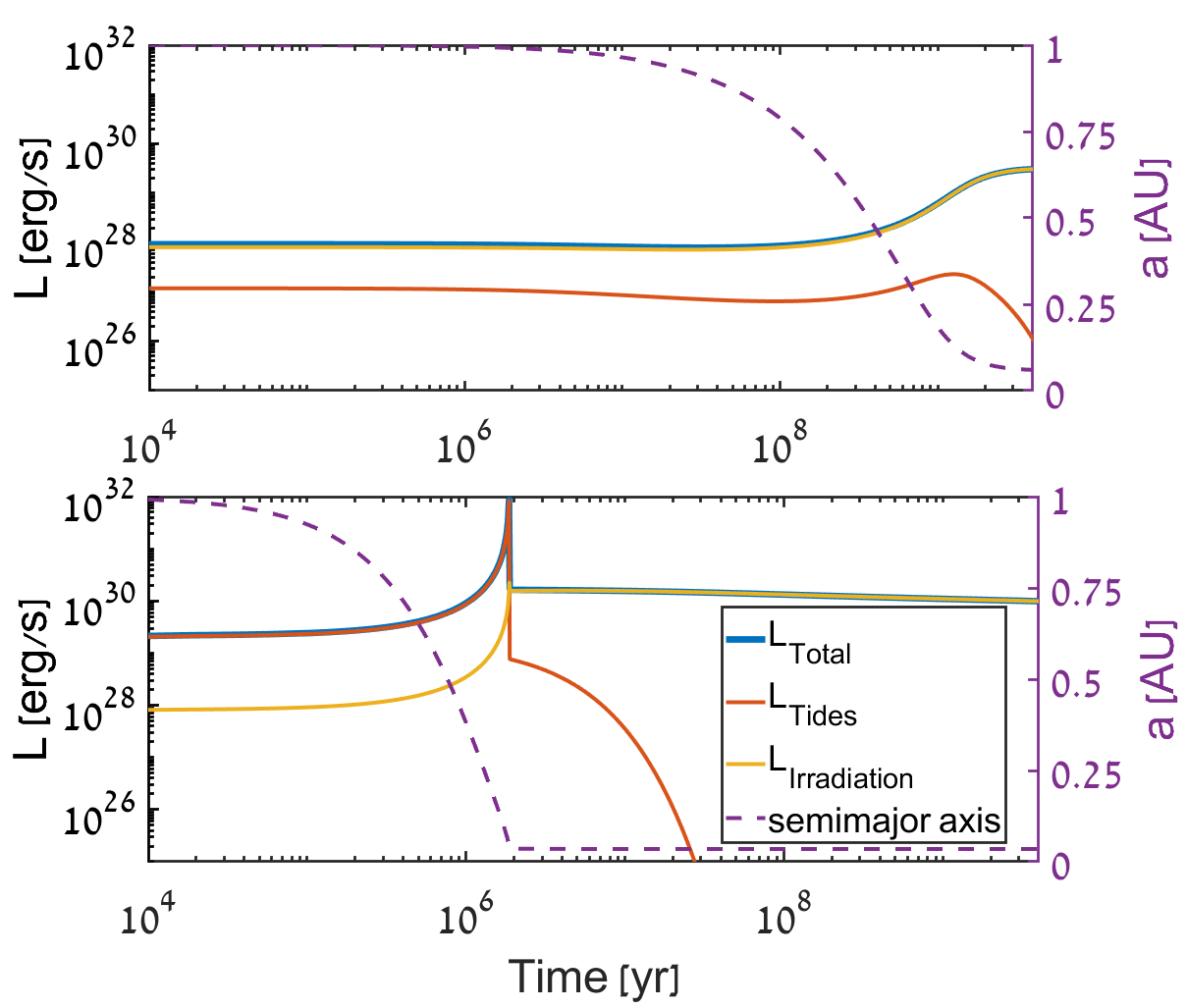}
\caption{The calculated external heat sources along the migration a $1 \rm M_J$ planet with an initial radius of $1.5\rm{R_J}$, initial semimajor axis of $1 \ \rm{AU}$ and initial eccentricity of $0.97$, under the influence of equilibrium tides (upper panel) and dynamical tides (lower panel) , simulations are presented in Fig. \ref{fig:compare_both_tides_1au0.97}.
 }
\label{fig:comb_L}
\end{figure}

Here and in paper I we considered the influence of two different external heat sources on the migration of HJs\&WJs candidates - irradiation from the host star 
and the energy from the tides acting on the planet. Our approach allows the inclusion of any external energy source and couple its effects to both the thermal and  dynamical processes. In Fig. \ref{fig:weak_extras} and Fig. \ref{fig:dynamical_extras} we demonstrate the differences in the migration of a gas giant with both tide models described in Sec. \ref{sec: high eccentricity tidal migration}, when including the different combination of heat sources in the thermal evolution of the planet. Both figures show the importance of the irradiation energy to achieve the observed effective temperature range of such planets. The effect of the tidal heating in these two cases is very minor in terms of the final properties of the planet, when the migration terminates. However, as was stated in Sec. \ref{subsec: dynamical tides}, the exact deposition of the dynamical tides inside the planet is unknown, in edition to its efficiency (i.e $f_\text{dyn}$) and the exact transition to equilibrium tides ($e_\text{trans}$). Therefore, the effect of the dynamical tides can be greater if the efficiency parameter is larger, as well as when the transition to weak tides occurs at a lower eccentricity. In the upper panel of Fig. \ref{fig:comb_L} we see that the radiation is indeed the most dominant along the entire evolution. On the other hand, the lower panel of same figure \ref{fig:comb_L} shows an example where the dynamical tides play the dominant role down to the transition point at $e_\text{trans}=0.2$. In this case, we used a value of $f_\text{dyn}=0.1$, while a larger value  would increase this effect, as can be seen in Fig. \ref{fig:dynamical_1au0.97_diff_f_etrans}.

We emphasize that there are many uncertainties regarding the deposition of heat by dynamical tides, which generally affect any eccentric migration model and the structure of tidal-migrating planets. We discuss the possible implications of the different assumptions on the HJs\&WJs population in paper I. However, in depth study of the exact behaviour of dynamical tides and their working is beyond the scope of this paper, where we consider several models and bracket their general implications.

\section{Discussion}
\label{Discussion}
\subsection{Inflated Hot Jupiters and heat transfer}
Although our study focuses on the early evolution of migrating Jupiters at which time they still retain large inflated radii following their initial formation, observations show the existence of at least some older inflated HJs, even at Gyr timescales. The abundance of such inflated HJs was suggested to indicate that an external source of deposited energy is required in order to keep HJs at an inflated phase or to re-inflate them after they already contracted \citep{GuillotShowman2002,Baraffe2010,ThorngrenFortney2018}.  
Several external energy sources and/or processes that conduct heat from the outer layers of the planet to the interior part (hence keeping the planets hotter) were suggested as a solution to the inflation \citep{GinzburgSari2015}. These include tidal heating \citep{Bodenheimer2001}, Ohmic heating \citep{BatyginStevenson2010} and irradiation from the star \citep{Burrows2003}. However, there is still no consensus on the origin of the population of such old inflated HJs.  Nonetheless, since observed WJs are usually not inflated \citep{MillerFortney2011}, this energy should potentially relate to the orbital separation from the host star, or to the migration timescales, which are correlated to the orbital energy and angular momenta of the migrating planet.
Depending on the energy source, its duration and its strength, it could potentially affect the migration process and shorten it.
More generally, if other processes exist that keep planets inflated, i.e. leading to even longer contraction timescales, our suggested inflated eccentric-migration should be even more efficient than already suggested by our results. 

\subsection{Internal distribution of energy}
\label{subsec:heat_tranfer}
\begin{figure*}
\centering
\includegraphics[width=0.87\linewidth]{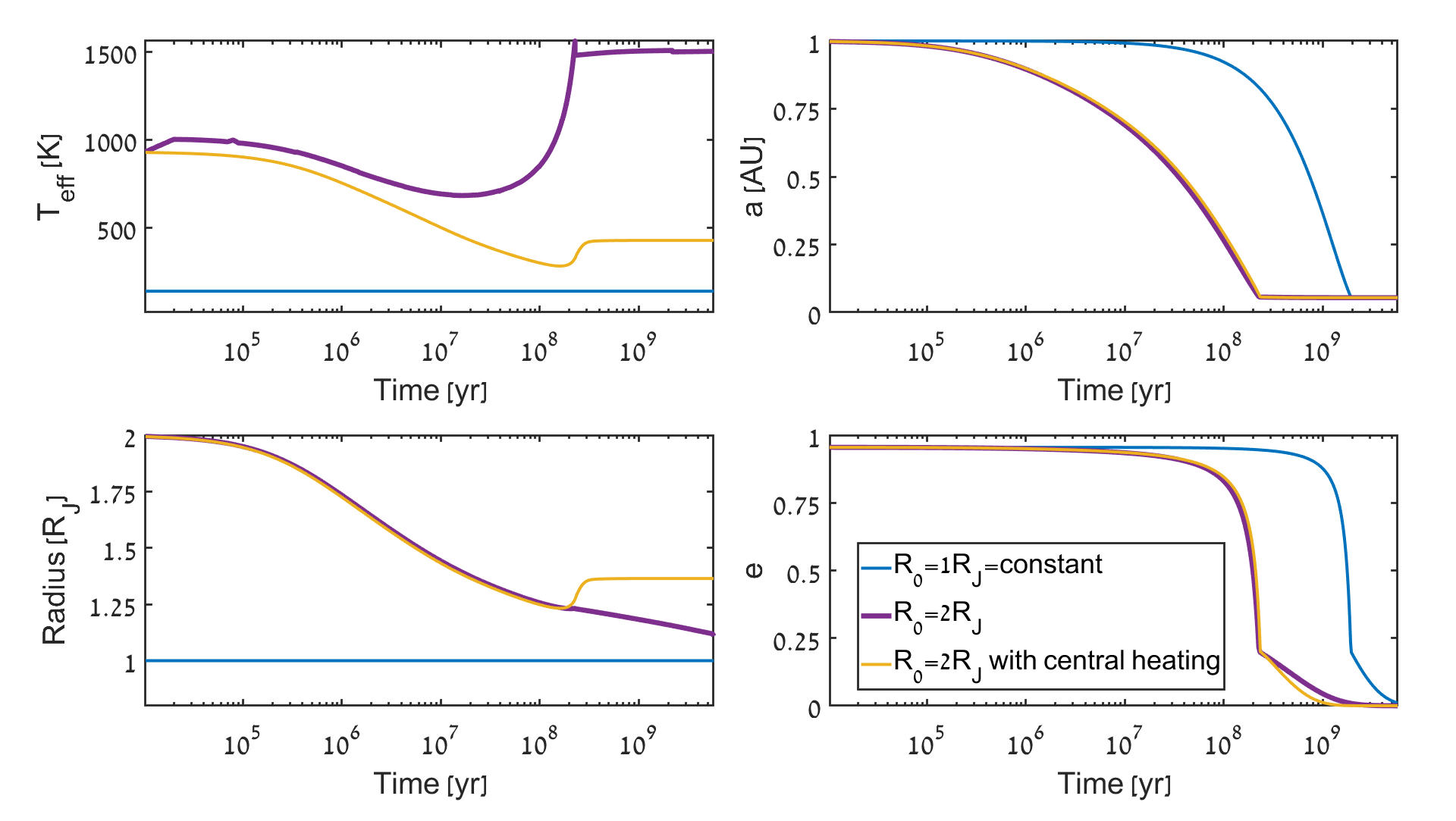}
\caption{Comparison of the migration of a constant ($R_0 = 1R_J$) radius, Jupiter mass planet (blue), and the migration of a similar planet which is also affected by dynamical tides and irradiation (purple) with a constant $R_0 = 1R_J$ (blue).
The injection of external heat is distributed according to eq. \ref{eq:extra_heat_distribution}. Yellow line corresponds to the case where only 1 \% of the external heat is injected, but it is now distributed around the center of the planet, i.e - at $r_\text{ext} = 0$. In all simulations $a_0 = 1 \rm AU$, $e_0 = 0.955$.
}
\label{fig:dynamical_with_dep}
\end{figure*}
When considering tidal heating and irradiation flux around the migrating planet's photosphere, we find that due to the efficient radiation of this energy the effect on the migration is mostly negligible (though is does  determine the planet's effective surface temperature). 

However, as can be seen in Fig. \ref{fig:dynamical_with_dep}, if the energy is deposited at a deeper region (when using smaller $r_\text{ext}$ in eq. \ref{eq:extra_heat_distribution}), the planet may slow its contraction or even re-inflate, thus its migration will be further accelerated. One can see that even a low efficiency of heat conductance to the center of the planet of only 1 \% of the energy is distributed around the center of the planet, when multiplying the expression of $\frac{dL_\text{ext}}{dm}$ in eq. \ref{eq:extra_heat_distribution} by $0.01$, the planet's radius can increase and affect the migration time. We note that the strength of the dissipation of irradiation energy on the planet, slightly varies due to the change in the mass distribution of the model and therefore change of $\sigma_\text{ext}$ and the result of the integral in eq. \ref{eq:extra_heat_distribution}.

\subsection{Formation of Warm Jupiters}
\label{subsec:WJs_discussion}
\begin{figure*}
\centering
\includegraphics[width=0.87\linewidth]{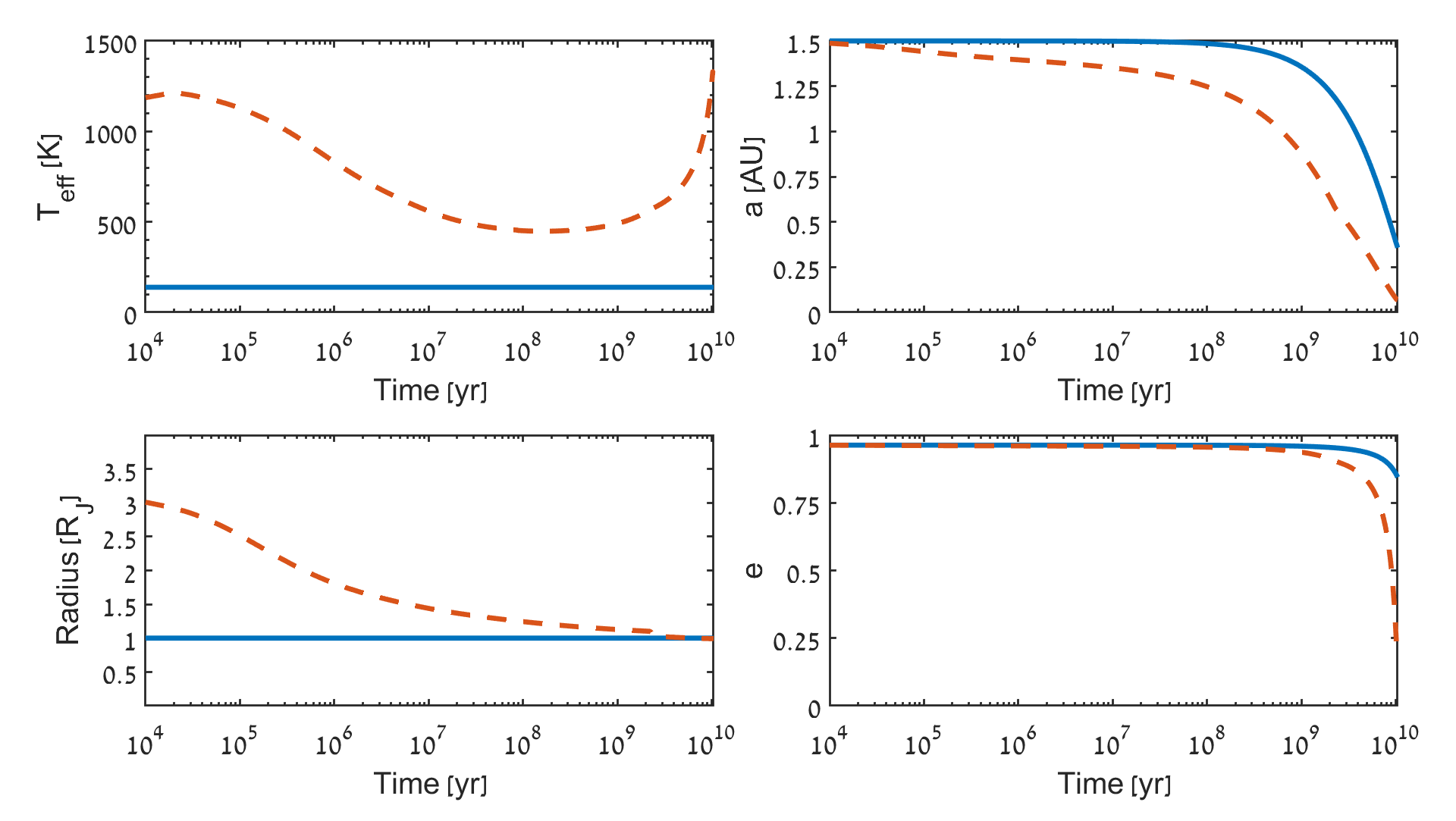}
\caption{The thermal and orbital evolution of hot and warm Jupiter candidates migrating due to dynamical tides, initialized with semimajor axis of $1.5 \ \rm{AU}$, and initial eccentricity of $0.963$. The blue line shows the dynamical evolution of a constant $R_0 = 1 \ R_J$,  whereas in orange is the evolution (thermal and dynamical) of $R_0=3 \ R_J$ affect by both irradiation and tidal energies. The initial $3 \ \rm{R_J}$ model reached an orbital period of $\sim 6.4 \ \rm{days}$ after a Hubble time, and became a HJ, and the constant $1 \rm{R_J}$ finalized at an orbit of $24.2 \ \rm{days}$, in the WJs regime.}
\label{fig:WJ}
\end{figure*}

Inflated eccentric migration enhances the migration rate such that planets that could not become HJ/WJs when considering an initial and constant $1 \ R_J$ radii, migrate more efficiently and now become WJs. Furthermore, inflated WJs, given the same initial conditions, would be less eccentric since they proceed faster in their migration; some of the expected WJs from the $1 \ R_J$ case will turn out to be HJs, since their inflated migration fastened to enable that.
Fig. \ref{fig:WJ} shows the evolution of two models of migrating WJs candidates during a Hubble time, where one outcome can be considered as a WJ (initialized with $R=1R_J$), and the other already as a HJ (initialized with $R=3R_J$). As both migration cases have not yet terminated at a Hubble time, one can deduce that ongoing star formation will enlarge the fraction of eccentric WJs, which are effectively on their way to become HJs on longer timescales. The fractions of WJs decay with time and the fraction of HJs increases, as WJ candidates end as HJ, if their migration is efficient enough. Therefore, star formation gives rise to increment in the fraction of WJs, on account of the fraction of HJs. 

\section{Summary}
\label{sec:Summary}
In this paper, together with paper I, we proposed a new efficient model for the formation of hot and warm Jupiters, by considering the radial/thermal evolution of the originally inflated planet along its migration. Here we used \texttt{AMUSE} \citep{Portegies2009AMUSE} to couple numerically the dynamical evolution of such planets according to different tidal models, with their internal evolution along the migration process (using \texttt{MESA} \citealt{PaxtonMESA2011}). 

Here and in paper I we showed that inflated eccentric migration process efficiently accelerates the migration of such gas-planets, compared with eccentric migration models where the thermal evolution of the planets is not considered. Initially inflated planets, and planets re-inflated due to tidal and/or radiative heating experience stronger tides, allowing for planets initialized at larger separations to migrate inwards, and induce higher rates of tidal disruptions of gas-giants. 

We find that the energy deposited by tides is mostly negligible in the equilibrium tides regime (weak tides), when deposited close to the planet's surface. Tidal heating can be important and even lead to planetary inflation if highly efficient dynamical tides are considered ($f_\text{dyn}>0.1$). In addition, efficient heat transfer from the outer regions of the planets where radiative and/or tidal heating is deposited to the central parts also give rise to significant thermal evolution and possible inflation of planets during their migration, even when only weak tides, or less efficient dynamical tides are considered. As the planets re-inflate, radii of HJs may become larger, but the number of disruptions may increase (see Paper-I for a further discussion). Identifying the exact processes and efficiencies of heat transfer in gas-giants is therefore critical for our understanding of their dynamical evolution and the formation of HJs and WJs. However, this is out of the scope of this paper, and we leave it to future works.

Our numerical and analytical approaches complement each other, and both can account for additional types of dynamical processes and other types of external energies.
   Our numerical model can be used to simulate the detailed evolution of stellar multiples where one can use the coupled internal evolution part on more than one component. The  good agreement between the numerical model presented here and the semi-analytic models presented in paper I supports the use of the latter, and the analytic model can well reproduce all of the numerical results. 
   
  The more computationally efficient semi-analytical could then be used to study the large parameter space of the HJ/WJ populations, as described in paper I.

\section{Software and third party data repository citations} \label{sec:cite}
Our numerical code can be found under the public repository- \url{https://github.com/hilaglanz/InflatedEccenricMigration}. 
Here we used \texttt{AMUSE} version 13.2.1 with self contribution as described, combined with \texttt{MESA} version 2208.
\software{AMUSE \citep{Portegies2009AMUSE},  
          MESA \citep{PaxtonMESA2011, PaxtonMESA2013}, 
          }
          
\begin{acknowledgments}
    We gratefully acknowledge helpful discussions with
    Sivan Ginzburg, Thaddeus D. Komacek, Michelle Vick, Nicholas C. Stone and Eden Saig. MR acknowledges the generous support of Azrieli fellowship.
\end{acknowledgments}

\bibliographystyle{aasjournal}

\end{document}